\documentclass[12pt]{article}           \long\def\del#1\enddel{}
\usepackage{latexsym,cite,epic}

\let\3=\ss \catcode`\"=\active \let"=\"	\parindent=20pt	\parskip=6pt

\newfont\blackboard{msbm10 scaled 1200}   %\font\blackboard=msbm10
\font\blackboards=msbm7		\font\blackboardss=msbm5
\newfam\black                   \scriptfont\black=\blackboards
\textfont\black=\blackboard     \scriptscriptfont\black=\blackboardss
\def\blackb#1{{\fam\black\relax#1}}	  
 \def\IZ{{\blackb Z}} \def\IQ{{\blackb Q}}  \let\BQ=\IQ
  \def\IP{{\blackb P}}  \let\BZ=\IZ

   \let\d=\delta \let\e=\varepsilon
    
\let\l=\lambda \let\m=\mu \let\n=\nu  \let\p=\pi

  \let\G=\Gamma \let\D=\Delta
\def\0{\over }    \def\1{\vec }   \def\2{{1\over2}} \def\3{{\ss}}
\def\4{{1\over4}} \def\5{\overline }   \def\6{\partial } \def\7#1{{#1}\llap{/}}
\def\8#1{{\textstyle{#1}}}        \def\9#1{{\bf {#1}}}

\def\<{\langle } \def\>{\rangle }  
 
\def\({\left( } \def \){\right) }
 
\let\and=\wedge  

\def\mao#1{\mathop{\rm {#1}}\nolimits}  \def\mod{\mao{mod}}

\def\beq{\begin{equation}} \def\eeq{\end{equation}} \def\eeql#1{\label{#1}\eeq}
\def\bea{\begin{eqnarray}} \def\eea{\end{eqnarray}} 
\def\beqnn#1\eeq{\[#1\]}   \let\nn=\nonumber
\def\fnote#1#2{\begingroup\def\thefootnote{#1}\footnote{#2}
           \addtocounter{footnote}{-1}\endgroup}

\def\plb#1 #2 {Phys. Lett. {\bf B#1} #2 }
\def\phr#1 #2 {Phys. Rep. {\bf  #1} #2 } 
\def\npb#1 #2 {Nucl. Phys. {\bf B#1} #2 }
\def\aph#1 #2 {Ann. Phys. {\bf #1} #2 }  
\def\jmp#1 #2 {J. Math. Phys. {\bf #1} #2 }
\def\prd#1 #2 {Phys. Rev. {\bf D#1} #2 }
\def\prl#1 #2 {Phys. Rev. Lett. {\bf #1} #2 }
\def\rmp#1 #2 {Rev. Mod. Phys.  {\bf #1} #2 }
\def\zpc#1 #2 {Z. Phys. {\bf #1C} #2 }
\def\cmp#1 #2 {Commun. Math. Phys. {\bf #1} #2 }
\def\cqg#1 #2 {Class.Quant.Grav. {\bf #1} #2 }
\def\mpl#1 #2 {Mod. Phys. Lett. {\bf A#1} #2 }
\def\ijmp#1 #2 {Int. J. Mod. Phys. {\bf A#1} #2 }

\def\ipb{\5{\hbox{\bf 1}}} \def\ip{\hbox{\bf 1}}  \def\Q{Q}

\def\ipo{\hbox{\bf 0}}	\def\fullA{\hbox{$A^I{}_J$}}

\def\BP{\begin{picture}} \def\EP{\end{picture}}		%% --> PICTURE macros
\def\putlin#1,#2,#3,#4,#5){\put#1,#2){\line(#3,#4){#5}}} %\putlin(x,y,dx,dy,l)
\def\putvec#1,#2,#3,#4,#5){\put#1,#2){\vector(#3,#4){#5}}}
\def\putc#1)#2{\put#1){\makebox(0,0)[cc]{#2}}}
\def\BC{\begin{center}}  \def\EC{\end{center}}

\def\pmb#1{\setbox0=\hbox{${#1}$}   \kern-.025em\copy0\kern-\wd0
      \kern.05em\copy0\kern-\wd0     \kern-.025em\raise.0433em\box0 }

\def\putdot#1){\putc#1){\circle*2}}	\def\putnum#1)#2{\putc#1){\pmb{_#2}}}

\long\def\new#1\endnew{{\bf #1}}

\newfont{\XLbf}{cmbx10 scaled 2800}	\newfont{\XL}{cmr10 scaled 2600}

\begin{document}	\def\wien{TUW--95/26}			% \draft

{\hfill hep-th/9512204   \vskip1pt \hfill\wien }
\vskip 19mm
\centerline{\XL		On the classification of reflexive polyhedra}
\begin{center} \vskip 12mm
       Maximilian KREUZER\fnote{*}{e-mail: kreuzer@tph16.tuwien.ac.at}
		              and
       Harald SKARKE\fnote{\#}{e-mail: skarke@tph16.tuwien.ac.at}
\vskip 5mm
       Institut f"ur Theoretische Physik, Technische Universit"at Wien\\
       Wiedner Hauptstra\3e 8--10, A-1040 Wien, AUSTRIA

\vfill                        {\bf ABSTRACT }
\end{center}

Reflexive polyhedra encode the combinatorial data for mirror pairs of
Calabi-Yau hypersurfaces in toric varieties. We investigate the geometrical
structures of circumscribed polytopes with a minimal number of facets
and of inscribed polytopes with a minimal number of vertices.
These objects, which constrain reflexive pairs of polyhedra from the interior
and the exterior, can be described in terms of certain non-negative integral
matrices. A major tool in the classification of these matrices is the
existence of a pair of weight systems, indicating a relation to weighted
projective spaces. This is the corner stone for an algorithm for the
construction of all dual pairs of reflexive polyhedra that we expect to be
efficient enough for an enumerative classification in up to 4 dimensions,
which is the relevant case for Calabi-Yau compactifications in string theory.

\vfill \noindent \wien\\[5pt] 
December 1995, revised version November 1996 \vspace*{9mm}
\thispagestyle{empty} 

\newpage
\pagestyle{plain}

%\newpage

\section{Introduction}

In the framework of toric geometry, it is possible to encode properties of
algebraic varieties in terms of fans or polyhedra defined on integer lattices.
In particular, it has been
shown by Batyrev that the Calabi--Yau condition for hypersurfaces
of toric varieties is equivalent to reflexivity of the underlying polyhedron
\cite{ba94}. Moreover, the duality of reflexive polyhedra corresponds to
the mirror symmetry of the resulting class of Calabi--Yau manifolds (see for
example \cite{as94,mo95,ca94,ho95} and references therein).
This is the main motivation for the interest  in a classification of
4-dimensional reflexive polyhedra in the context of string theory.

It is known that the total number of reflexive polyhedra is finite in any
given dimension, because various bounds on the volume and the number of points
have been derived as a function of the dimension and the number of interior
lattice points \cite{ba82,he83,bo92}.
The case of $n=2$ dimensions is the easiest because all polygons with one
interior point are reflexive (this is no longer true for $n>2$).
There are 16 such polygons, which were constructed in
\cite{ba85,ko90} (we will rederive this
result in the last section to illustrate the application of our tools).
In 4 dimensions we expect  at least  some $10^4$ reflexive pairs and the
known bounds
for general lattice polytopes \cite{ba82,he83,bo92} are not very useful for
explicit constructions.
What we need is an efficient algorithm, which probably should
rely on reflexivity in an essential way. It is the purpose of the present
paper to provide such an algorithm.

Our approach is partly motivated by experience with transversal
polynomials in weighted projective spaces \cite{fl89,cqf,nms,kl94}
and by the orbifold construction of mirror pairs \cite{gr90,be93,aas,mmi},
but this will become clear only at a later stage. The basic strategy is to
find minimal integral polytopes $M$ that are spanned by vertices of $\D$ and
that still have \ipo\ in the interior (the generic case is a simplex).
By duality,  $M^*$ bounds $\D^*$ and its facets carry facets of $\D^*$.
If we have minimal polytopes $M$ and $\5M$ for $\D$ and $\D^*$,
respectively, then the pairing matrix of the respective vertices turns out
to be strongly constrained. 
\del
In fact, there is a simple algorithm for the
iterative computation of all candidate matrices $A^i{}_j$.
\enddel
Such a matrix encodes the structures of $M$ and $\5M^*$, which bound
$\D$ from the interior and the exterior.

The final step in the classification is the reconstruction of the complete
pairing matrix %$\fullA$ 
of vertices of $\D$ and $\D^*$.
\del
The condition that $\fullA$ contains $A^i{}_j$ as a submatrix strongly
constrains the additonal vertices that may be added. 
\enddel
The pairings of all vertices characterize the reflexive pair up to a finite 
number of possible choices of dual pairs of lattices.
In the simplex case the barycentric coordinates of the interior point
correspond to the weights in the context of weighted projective spaces.
Indeed, the authors of \cite{ca95} tried to interpret toric Calabi-Yau
manifolds as non-transverse hypersurfaces in weighted $\IP^4$.
Our results imply that, even without transversality, only a finite number of
weight systems makes sense in the toric context.
Moreover, the large ambiguity in the generalized transposition rule of
\cite{ca95}
is constrained by our rules for the selection of vertices, which may be
regarded as rules about which transpositions make sense.

In section 2 we give some basic definitions and deduce
geometrical properties of minimal polytopes.
In general we may need a number of lower dimensional simplices containing
\ipo\ in the interior to span a neigborhood of $\ipo$. Then we have
several weight systems and the toric variety can only be related to sort of
a (non-direct) product of weighted 	% projective
spaces.
In section 3 we discuss the properties of (minimal) pairing matrices and
the relations among pairings in higher-dimensional lattices that we use
to embed a reflexive pair. 
We illustrate our concepts using an example of a 4-dimensional polyhedron 
that was analysed in the toric context in \cite{ho95}. 
This completes the setup that we need in section 4
to state the classification algorithm and to prove its finiteness. 
As an illustration we rederive the 2-dimensional case.

\section{Reflexive polyhedra and minimal polytopes}

We first recall some elementary definitions about polytopes \cite{ZI95}.
A rational polyhedron is an intersection of finitely many halfspaces
$\{x\in\IQ^n: a_ix^i\ge b \}$ with $a_i,b\in\IZ$. A polytope is a bounded
polyhedron or, equivalently, the convex hull of a finite number of points.
A lattice (or integral) polytope  is a polytope whose vertices
belong to some lattice $\G\cong\IZ^n$. 
We will identify $\IQ^n$ with the rational extension $\G_\IQ$ of $\G$,
i.e. $\IQ^n\cong\G_\IQ=\G\otimes_\IZ\IQ$.
The distance of a lattice point $x\in\G$ to a lattice hyperplane
$H(a_i,b)=\{x\in\G_\IQ:a_ix^i=b\}$ where the integers $a_i$ have greatest 
common divisor 1, is defined by $d(H,x):=|a_ix^i-b|$.
This number is 1 plus the number of lattice hyperplanes between $x$ and $H$.
These definitions are invariant under changes of the lattice
basis, so we can write $\<a,x\>$ instead of $a_ix^i$ whenever we
do not want to refer to a specific basis.
Then the condition that the $a_i$ have no common divisor means that
$a\in\G^*$ is primitive.

A reflexive polyhedron $\D$ is a polytope
with one interior point $P$ whose bounding hyperplanes are all at distance 1
from $P$. If an arbitrary convex set $\D$ contains the origin in its
interior we define the dual (or polar) set
\beq
	\D^*:=\{y\in \G^*_\IQ:\<y,x\>\ge-1\:\forall x\in\D\}.
\eeq
Assuming that the interior point $P=\ipo$ is the origin it is easy to see that
a polytope $\D$ is reflexive if and only if $\D^*$ is integral, i.e. if all
vertices of $\D^*$ belong to $\G^*$.

Consider an $n$-dimensional reflexive pair of polyhedra
$\D$ and $\D^*$ defined on lattices $\G$ and $\G^*$.
For each of these polyhedra we choose
a set of $k$ ($\5k$) hyperplanes $H_i$, $i=1,\cdots, k$
($\5H^j$, $j=1,\cdots, \5k$) carrying facets in such a way that
these hyperplanes define a bounded convex body $Q$ ($\5Q$) containing
the original polyhedron and that $k$ and $\5k$ are minimal.
We define a redundant
coordinate system where the $i^{th}$ coordinate of a point is given by its 
integer distance to $H_i$ (nonnegative on the side of the polyhedron).
This is just the degree of the homogeneous coordinate \cite{cox}
corresponding to $H_i$ in the monomial determined by the point.  
Note that the vertices of $Q$ and $\5Q$ need not have integer coordinates.
All coordinates of the interior points  are equal to $1$,
each coordinate of any point of a polyhedron is nonnegative.
Whenever we use this sort of coordinate system we will label the interior
points by $\ip$ and $\ipb$.
Note that $\ip$ is the only integer point in the interior of $Q$:
For all other points  one coordinate  must be smaller than one so that they
belong to some $H_i$. We have thus shown that any such polytope $Q$
has all lattice points of $\D$, except for $\ip$, at its boundary.

The duals of these hyperplanes are two collections of $\5k$ ($k$)
vertices $V_j$ ($\5V^i$) spanning polyhedra $M=\5Q^*$ and $\5M=Q^*$
that contain the interior
points of $\D$ ($\D^*$). $M$ and $\5M$ are minimal in the sense
that there are no collections of less than $\5k$ ($k$) vertices
of $\D$ ($\D^*$) containing $\ip$ ($\ipb$) in the interior.

Let us first obtain some information on the general structure of
minimal polytopes.
Here we will not use the affine structure
(labelling the interior point by $\ip$),
 but instead we will use a linear structure, calling the
interior point $\ipo$ and identifying vertices $V$ with vectors.
Then the fact that $M$ has $\ipo$ in its interior is equivalent to
the fact that any point in $\BQ^n$ can be written as a nonnegative linear
combination of vertices.
Considering all triangulations where every simplex contains a specific
vertex $V$ of $M$, we see that there is at least one simplex of dimension
$n$ with this vertex containing $\ipo$, i.e. $\ipo$ lies in the interior of
this simplex or one of its simplicial faces containing $V$.
So we have a collection of vertices and a collection of subsets
of this set of vertices defining lower dimensional simplices with $\ipo$
in their interiors (we will call such simplices ``good simplices''), in
such a way that each vertex belongs to at least one good simplex.
Now we note that if we have a collection of good simplices, then $\ipo$ is
also in the interior of the polytope spanned by all the vertices of these
simplices (of course, ``interior'' here means interior w.r.t. the
linear subspace spanned by these vertices).
\\[4pt]
{\bf Lemma 1:} A minimal polytope $M=\mao{Convex Hull}\{V_1,\cdots,V_{\5k}\}$
in $\BQ^n$ is either a simplex or contains an $n'$-dimensional minimal
polytope $M':=\mao{Convex Hull}\{V_1,\cdots,V_{\5k'}\}$ and a good simplex
$S:=\mao{Convex Hull}(R\cup\{V_{\5k'+1},\cdots,V_{\5k}\})$ with
$R\subset\{V_1,\cdots,V_{\5k'}\}$ such that $\5k-\5k'=n-n'+1$ and
$\mao{dim}S\le n'$.\\
{\it Proof:}
If $M$ is a simplex, there is nothing left to prove. Otherwise,
consider the set of all good simplices consisting of vertices of
$M$. Any subset of this set will define a lower dimensional minimal polytope.
Among these, take one (call it $M'$) with the maximal dimension $n'$
smaller than $n$.
$\BQ^n$ factorizes into $\BQ^{n'}$ and $\BQ^n/\BQ^{n'}\cong\BQ^{n-n'}$
(equivalence classes in $\BQ^n$). The remaining vertices define a
polytope $M_{n-n'}$ in $\BQ^n/\BQ^{n'}$. If $M_{n-n'}$ were not a simplex,
it would contain a simplex of dimension smaller than $n-n'$ which would
define, together with the vertices of $M'$, a minimal polytope
of dimension $s$ with $n'<s<n$, in contradiction with our assumption.
Therefore $M_{n-n'}$ is a simplex. Because of minimality of $M$,
each of the $n-n'+1$ vertices of $M_{n-n'}$ can have only one representative
in $\BQ^n$, implying $\5k-\5k'=n-n'+1$.
The equivalence class of $\ipo$ can be described uniquely as a positive
linear combination of these vertices.
This linear combination
defines a vector in $\BQ^{n'}$, which can be written as a negative linear
combination of $\le n'$ linearly independent vertices of $M'$.
These vertices, together with those of $M_{n-n'}$, form the simplex $S$.
By the maximality assumption about $M'$, dim$S$ cannot exceed dim$M'$.
\hfill$\Box$\\[4pt]
{\bf Corollary 1:}
A minimal polytope $M=\mao{Convex Hull}\{V_1,\cdots,V_{\5k}\}$ in $\BQ^n$
allows a structure \\
$\{V_j\}=\{V_1,\cdots,V_{\5k_1},V_{\5k_1+1},\cdots,V_{\5k_2},\cdots,
V_{\5k_{\l-1}+1},\cdots,V_{\5k_\l}\}$ with the following properties:\\
(a) $M_\m:=\mao{Convex Hull}\{V_1,\cdots,V_{\5k_\m}\}$ is a
$(\5k_\m-\m)$ -- dimensional minimal polytope, $M_\l=M$.\\
(b) For each $\m$, there is a subset $R_\m$ of $\{V_1,\cdots,V_{\5k_{\m-1}}\}$
such that
$S_\m:=\mao{Convex Hull}(R_\m\cup\{V_{\5k_{\m-1}+1},\cdots,V_{\5k_\m}\})$
defines a simplex with $\mao{dim}S_\m\le\mao{dim}M_{\m-1}$ for $\m>1$.\\
{\it Proof:}
If $M$ is a simplex, $\l=1$ and $\5k=\5k_1=n+1$. 
Otherwise one can proceed inductively using lemma 1.
\hfill$\Box$\\[4pt]
{\bf Corollary 2:} $n+1\le \5k\le2n$.\\[2pt]
{\it Proof:} If $M$ is a simplex, the lower bound is satisfied.
Otherwise, $\5k=\5k'+n-n'+1$ and induction gives
$\5k\in%\{n'+1+n-n'+1,\cdots, 2n'+n-n'+1\}=
\{n+2,\cdots, n'+n+1\}\subset\{n+1,\cdots, 2n\}$.
\hfill$\Box$\\[4pt]
{\bf Lemma 2:} Denote by $\{S_\l\}$ a set of good simplices spanning $M$. Then
$S_\m-\bigcup_{\n\ne\m}S_\n$ never contains exactly one point.\\[2pt]
{\it Proof:} A simplex with $\ipo$ in its interior contains line
segments $VV'$ with $V'=-\e V$, where $\e$ is a positive number.
If a simplex $S=\mao{Convex Hull}\{V_1,\cdots, V_{s+1}\}$ has all of its
vertices except one ($V_{s+1}$) in common with
other simplices, then all points in the linear span of
$S$ are nonnegative linear combinations of the $V_j$ and the $-\e_jV_j$
with $j\le s$, thus showing that $V_{s+1}$ violates the minimality of $M$.
\hfill$\Box$\\[4pt]
{\bf Example:} $n=5$, $M=\mao{Convex Hull}\{V_1,\cdots,V_8\}$ with
\bea V_1=(1,1,0,0,0)^T,\;V_2=(1,-1,0,0,0)^T,\;
     V_3=(-1,0,1,0,0)^T,\;V_4=(-1,0,-1,0,0)^T,\nn\\
     V_5=(-1,0,0,1,0)^T,\;V_6=(-1,0,0,-1,0)^T,\;
     V_7=(1,0,0,0,1)^T,\;V_8=(1,0,0,0,-1)^T.    \eea
$M$ contains the good simplices $S_{1234}=V_1V_2V_3V_4$ (in the
$x_1x_2x_3$--plane),
$S_{1256}$ (in the $x_1x_2x_4$--plane),
$S_{3478}$ (in the $x_1x_3x_5$--plane),
$S_{5678}$ (in the $x_1x_4x_5$--plane) and the 4-dimensional
minimal polytopes
$M_{123456}$, $M_{123478}$, $M_{125678}$, $M_{345678}$.
Each of these 4-dimensional minimal polytopes spans a hyperplane
of codimension 1. In order to span $\BQ^5$, we need two additional points
which may belong to one of two possible simplices. For example, if
we choose $M=M_{123456}$, then we require $V_7$ and $V_8$ and may choose
$S= S_{3478}$ or $S= S_{5678}$. $M_{123456}$ again has to fulfill lemma 1,
and indeed it contains the two simplices $S_{1234}$ and $S_{1256}$.
%=V_1V_2V_3V_4V_5V_6$ (in the $x_1x_2x_3x_4$--plane).
The structure of corollary 1 can be realised, for example, by
$S_1=M_1=S_{1234}$, $S_2=S_{1256}$ (implying $M_2=M_{123456}$), and
$S_3=S_{3478}$. Note that $S_{5678}$ does not occur in this structure and
that $S_1-(S_2\cup S_3)$ is empty (compare with lemma 2).

\section{Pairing matrices}

Let the elements ${A^i}_j$ of the integer $k\times\5k$ matrix $A$ denote 
the $i^{th}$ coordinate of $V_j$ in the coordinate system defined by $Q$, i.e.
${A^i}_j=(V_j)^i$ is the distance of $V_j$ to the hyperplane $H_i$.
Because of reflexivity of $\D^*$ (all facets are at distance 1 from $\ipb$)
this is related to the pairing of $V_j$ with $\5V^i=H_i^*$ by
${A^i}_j=\<\5V^i,V_j\>_+$ with $ \<\;,\;\>_+:=\<\;,\;\>+1$, where
$\<\;,\;\>$ is the original lattice pairing.
The definition of the affine pairing $ \<\;,\;\>_+$ might seem awkward
at first sight, but it has two advantages: On the one hand,
it is nonnegative for any pairing between $\D^*$ and $\D$, and on the
other hand, we will see later that it is a natural linear pairing for
a higher dimensional pair of lattices into which we will embed $\G$ and $\G^*$.
By duality ${A^i}_j$ also denotes the $j^{th}$
coordinate of $\5V^i$ in the coordinate system defined by $\5Q$, i.e.
${A^i}_j=(\5V^i)_j$.
In other words, the columns of $A$ correspond to the vertices of $M$ whereas
the lines correspond to the vertices of $\5M$. We will label all points
of $\D$ by column vectors and all points of $\D^*$ by line vectors,
in particular $\ip=(1,\cdots,1)^T$ and $\ipb=(1,\cdots,1)$.

If $M$ and $\5M$ are simplices, then $A$ is an $(n+1)\times (n+1)$ matrix.
We denote by the ``weights'' $q_i$ and $\5q^j$ the barycentric
coordinates of $\ipb$ and $\ip$, respectively:
\beq  \ipb=\sum_i q_i\5V^i,\qquad \ip=\sum_j\5q^jV_j,\qquad
     \sum_i q_i=\sum_j\5q^j=1. \eeq
This implies
\beq
	\sum_i q_i {A^i}_j=1, ~~~~~~~ \sum_j \5q^j {A^i}_j=1, ~~~~~~~
        \sum_i q_i (V_j)^i=1, ~~~~~~~ \sum_j \5q^j (\5V^i)_j=1.
\eeql{qdef}
We can now give a new interpretation to our coordinate systems
as coordinates in
%$\G^{n+1}_\BQ$ and $\5\G^{n+1}_\BQ$. Consider the
$(n+1)$--dimensional lattices $\G^{n+1}\cong\BZ^{n+1}$
and $\5\G^{n+1}\cong\BZ^{n+1}$ and their rational extensions
$\G^{n+1}_\BQ\cong\BQ^{n+1}$ and $\5\G^{n+1}_\BQ\cong\BQ^{n+1}$.
Then the equation $\sum q_ix^i=1$ defines an $n$--dimensional affine
hyperplane $\G^n_\BQ$ spanned by the $V_j$, which obviously contains
$\ip$. Linear independence of the $V_j$,
i.e. of the columns of $A$, implies regularity of $A$.
We can invert Eqs. (\ref{qdef}) to get
\beq
     q_i=\sum_{j=0}^n(A^{-1})^j{}_i, ~~~~~~~ \5q^j=\sum_{i=0}^n(A^{-1})^j{}_i.
\eeq
Defining arbitrary point pairings $\<\;,\;\>_{n+1}$
between $\G^{n+1}_\BQ$ and $\5\G^{n+1}_\BQ$ by
\beq
      \< \5P,P\>_{n+1}:=\5P_k(A^{-1})^k{}_lP^l
\eeq
allows us to identify $\G^n_\BQ$ and $\5\G^n_\BQ$ as
\beq
     \G^n_\BQ\cong\{P\in \G^{n+1}_\BQ ~| ~~\<\ipb,P\>_{n+1}=1\}, ~~~~~~~~~~
     \5\G^n_\BQ\cong\{\5P\in \5\G^{n+1}_\BQ ~| ~~\<\5P,\ip\>_{n+1}=1\}.
\eeq
At this point it is easy to see the relation of our framework to the
orbifold mirror construction that works for minimal polynomials
in weighted projective spaces \cite{gr90,be93}: That construction relates
a monomial with exponent vector $\5W$ to a twist group element whose diagonal
action on the homogeneous coordinates is $\exp(2\p i\mao{diag}(\5W A^{-1}))$
\cite{mmi}. Even in our more general context the lines of $A^{-1}$ provide
the phases for generators of the phase symmetry group of the $n+1$
monomials whose exponents are the columns of the matrix $A$
(this does not mean, however, that all such symmetries can be used
for an orbifold construction, because transversality requires additional
monomials in the non-minimal case; we will soon give an example for how
this manifests itself in the context of toric geometry).

\noindent
{\bf Lemma 3:} Let $P\in\G^n_\BQ,\;\5P\in\5\G^n_\BQ$.
Then\\
(a) $\< \5P,P\>_{n+1}=\< \5P,P\>_+$.\\
(b) $\< \5P-\ipb,P-\ip\>_{n+1}=\< \5P,P\>$.\\[2pt]
{\it Proof:}
For vertices we have by definition
\beq \< \5V^i,V_j\>_{n+1}=(\5V^i)_k(A^{-1})^k{}_l(V_j)^l=
    A^i{}_k(A^{-1})^k{}_lA^l{}_j={A^i}_j=\< \5V^i,V_j\>_+\eeq
and
\beq \< \5V^i-\ipb,V_j-\ip\>_{n+1}=
%\< \5V^i,V_j\>_{n+1}-
%\<\ipb,V_j\>_{n+1}-\< \5V^i,\ip\>_{n+1}+\< \ipb,\ip\>_{n+1}=
\< \5V^i,V_j\>_+-1-1+1=\< \5V^i,V_j\>.
\eeq
For general $P$, $\5P$ (b) follows from linearity in $\G^n_\BQ$ and
$\5\G^n_\BQ$ and (a)
follows from (b) because
$\<\5P,\ip\>_{n+1}=\<\ipb,P\>_{n+1}=\<\ipb,\ip\>_{n+1}=1$.
\hfill	$\Box$

The first statement of this lemma shows us that $\<\;,\;\>_{n+1}$
is a natural extension of $\<\;,\;\>_+$ to $\G^{n+1}_\BQ\times\5\G^{n+1}_\BQ$.
We will use this fact to define $\<\;,\;\>_+$
in $\G^{n+1}_\BQ\times\5\G^{n+1}_\BQ$, thus showing that our
originally affine pairing is indeed a linear pairing in the higher
dimensional context.
Let us also define  the $n$-dimensional sublattices
$\G^n=\G^n_\BQ\cap\G^{n+1}$ and $\5\G^n=\5\G^n_\BQ\cap\5\G^{n+1}$
carrying $\D$ and $\D^*$, respectively.

\noindent
{\bf Corollary 3:} There is a natural identification
	$(\G^n)^*\cong \ipb+\mao{Span}\{\5V^i-\ipb\}\subseteq\5\G^n$.
\\[2pt]
{\it Proof:}
By the embedding of $\G^n$ into $\IZ^{n+1}$ an element of $(\G^n)^*$
becomes an equivalence class of points in the dual lattice
$\IZ^{n+1}$ modulo $\ipb$.
Since $(\5V^i)_k(A^{-1})^k{}_l=\d^i_l$ the vertices $\5V^i$ are representatives
of equivalence classes that generate $(\G^n)^*$.
Using the $\mod\ipb$ ambiguity we may
always choose a representative in $\ipb+\mao{Span}\{\5V^i-\ipb\}$
because $\<\ipb,P\>_{n+1}=1$.
\hfill	$\Box$

Given a pairing matrix $A$ for our simplices $M$ and $\5M$, let us see how 
we can obtain all corresponding dual pairs $\D$, $\D^*$:
First we choose some sublattice $\G\subseteq\G^n$ that contains $\ip$ and
all vectors $V_j$. 
The dual lattice $\G^*$ is a sublattice of $\5\G^n$, which
obviously contains the vectors $\5V^i$. Then
\beq
\Q=\{P\in \G^{n+1}_\BQ~|~\<\ipb,P\>_+=1 \and P^i\geq0~\forall i\}, ~~~~~
\5\Q=\{\5P\in \5\G^{n+1}_\BQ~|~\<\5P,\ip\>_+=1\and \5P_j\geq0~\forall j\}.
\eeq
\del
We can now choose a
dual pair of lattices $\G$ and $\G^*$ with
$\ip+\mao{Span}\{V_j-\ip\}\subseteq\G\subseteq\G^n$ and
$\ipb+\mao{Span}\{\5V^i-\ipb\}\subseteq\G^*\subseteq\5\G^n$.
\enddel
Defining the finite point sets $\G_+=\{P\in\G~|~P^i\ge0~\forall i\}$ and
$(\G^*)_+$ and their convex hulls $\D_{\rm max}$ and $\5\D_{\rm max}$, 
respectively, we may choose polyhedra $\D$ and $\5\D$ with 
$\{V_j\}\subseteq\D\subseteq\D_{\rm max}$ and 
$\{\5V^i\}\subseteq\5\D\subseteq\5\D_{\rm max}$ and check for duality.
In practice, the following algorithm will be far more efficient:
Calculate all points $P,\5P$ in $\G^n_+$ and $\5\G^n_+$ and the
corresponding pairing matrix (w.r.t. $\<\;,\;\>_+$), which may have
rational entries.
\del
In the case of simplices we can use $A^{-1}$ for the evaluation of the pairing
as described in the previous section; in the general case we may either work
with some basis of $\G$ or use a
regular $(n+1)\times(n+1)$ submatrix $\hat A$ of $A$ which can be used
in the same way as $A$ was used in the simplex case (note that $\hat A$
also defines simplices in $\G^n_\BQ$ and $\5\G^n_\BQ$, these do not have
$\ip$ and $\ipb$ in their interiors, however).
\enddel
Then we can create a list of possible vertices $V$ by noting that
any vertex is dual to a hyperplane, i.e. for any vertex $V$
there must be $n$ linearly independent points $\5P$ with $\<\5P,V\>_+=0$.
Creating a list of possible vertices $\5V$, we use the same argument,
working only with our list of possible vertices in $\G^n_+$.
This procedure may be iterated, reducing the respective lists in
each step. In particular we can drop a model whenever our original
vertices $V_j$ or $\5V^i$ don't show up in the resulting lists of
possible vertices. In a last step we may then choose subsets of these
lists, making sure that each coordinate hyperplane contains $n$
linearly independent vertices.
Choosing a particular point $P$ to be a vertex of $\D$ implies that
we can eliminate all points $\5P$ with rational or negative pairings
with $P$ from our list of candidates for vertices of $\D^*$.

\begin{figure}
{\tiny
\[
\hspace*{-11mm}
\pmatrix{25 & 1 & 0 & 0 & 1 & 0 & 0 & 2 & 2 & 3 & 3 & 3 & 4 & 4 & 5 & 5 & 6
    & 7 & 8 & 8 & 9 & 9 & 10 & 11 & 12 & 13 & 14 & 16 & 17 & 19 & 22 \cr
&&&&&&&&&&&&&&&&&&&&&&&&&&&&&& \cr 0 & 8
    & 0 & 1 & 0 & 3 & 4 & 2 & 5 & 0 & 2 & 3 & 0 & 7 & 1 & 4 & 1 & 6 & 0 & 3 &
   0 & 1 & 5 & 2 & 0 & 4 & 1 & 3 & 0 & 2 & 1 \cr
&&&&&&&&&&&&&&&&&&&&&&&&&&&&&& \cr 0 & 0 & 3 & 0 & 0 & 0 & 1 & 3
    & 0 & 0 & 0 & 1 & 1 & 0 & 3 & 0 & 0 & 0 & 3 & 0 & 0 & 1 & 0 & 0 & 1 & 0 &
   0 & 0 & 0 & 0 & 0 \cr
&&&&&&&&&&&&&&&&&&&&&&&&&&&&&& \cr 0 & 0 & 0 & 3 & 0 & 0 & 1 & 0 & 0 & 3 & 0 &
1 & 1 & 0
    & 0 & 0 & 0 & 0 & 0 & 0 & 0 & 1 & 0 & 0 & 1 & 0 & 0 & 0 & 0 & 0 & 0 \cr
&&&&&&&&&&&&&&&&&&&&&&&&&&&&&& \cr 0
    & 0 & 1 & 0 & 3 & 2 & 0 & 0 & 1 & 0 & 2 & 0 & 1 & 0 & 0 & 1 & 2 & 0 & 0 &
   1 & 2 & 0 & 0 & 1 & 0 & 0 & 1 & 0 & 1 & 0 & 0 \cr
&&&&&&&&&&&&&&&&&&&&&&&&&&&&&& \cr 0 & 0 & 0 & 1 & 3 & 2 & 0
    & -1 & 1 & 1 & 2 & 0 & 1 & 0 & -1 & 1 & 2 & 0 & -1 & 1 & 2 & 0 & 0 & 1 & 0
    & 0 & 1 & 0 & 1 & 0 & 0 \cr
&&&&&&&&&&&&&&&&&&&&&&&&&&&&&& \cr
0 & 0 & 1 & 2 & 0 & 0 & 1 & 1 & 0 & 2 & 0 & 1
    & 1 & 0 & 1 & 0 & 0 & 0 & 1 & 0 & 0 & 1 & 0 & 0 & 1 & 0 & 0 & 0 & 0 & 0 &
   0 \cr
&&&&&&&&&&&&&&&&&&&&&&&&&&&&&& \cr 0 & 0 & 2 & 1 & 0 & 0 & 1 & 2 & 0 & 1 & 0 &
1 & 1 & 0 & 2 & 0 & 0 & 0
    & 2 & 0 & 0 & 1 & 0 & 0 & 1 & 0 & 0 & 0 & 0 & 0 & 0 \cr
&&&&&&&&&&&&&&&&&&&&&&&&&&&&&& \cr 0 & 4 & 0 & 0 & 3
    & {7\over 2} & {3\over 2} & 0 & {7\over 2} & -{1\over 2} & 3 & 1 &
   {1\over 2} & {7\over 2} & -{1\over 2} & 3 & {5\over 2} & 3 & -1 &
   {5\over 2} & 2 & 0 & {5\over 2} & 2 & -{1\over 2} & 2 & {3\over 2} &
   {3\over 2} & 1 & 1 & {1\over 2} \cr
&&&&&&&&&&&&&&&&&&&&&&&&&&&&&& \cr 0 & 4 & 0 & 2 & 0 & {3\over 2} &
   {5\over 2} & 1 & {5\over 2} & {3\over 2} & 1 & 2 & {1\over 2} & {7\over 2}
    & {1\over 2} & 2 & {1\over 2} & 3 & 0 & {3\over 2} & 0 & 1 & {5\over 2} &
   1 & {1\over 2} & 2 & {1\over 2} & {3\over 2} & 0 & 1 & {1\over 2} \cr
&&&&&&&&&&&&&&&&&&&&&&&&&&&&&& \cr 0 & 4
    & 1 & 1 & 0 & {3\over 2} & {5\over 2} & 2 & {5\over 2} & {1\over 2} & 1 &
   2 & {1\over 2} & {7\over 2} & {3\over 2} & 2 & {1\over 2} & 3 & 1 &
   {3\over 2} & 0 & 1 & {5\over 2} & 1 & {1\over 2} & 2 & {1\over 2} &
   {3\over 2} & 0 & 1 & {1\over 2} \cr
&&&&&&&&&&&&&&&&&&&&&&&&&&&&&& \cr 0 & 4 & 2 & 0 & 0 & {3\over 2} &
   {5\over 2} & 3 & {5\over 2} & -{1\over 2} & 1 & 2 & {1\over 2} & {7\over 2}
    & {5\over 2} & 2 & {1\over 2} & 3 & 2 & {3\over 2} & 0 & 1 & {5\over 2} &
   1 & {1\over 2} & 2 & {1\over 2} & {3\over 2} & 0 & 1 & {1\over 2} \cr
&&&&&&&&&&&&&&&&&&&&&&&&&&&&&& \cr 0 & 8
    & 1 & 0 & 0 & 3 & 4 & 3 & 5 & -1 & 2 & 3 & 0 & 7 & 2 & 4 & 1 & 6 & 1 & 3
    & 0 & 1 & 5 & 2 & 0 & 4 & 1 & 3 & 0 & 2 & 1 \cr
&&&&&&&&&&&&&&&&&&&&&&&&&&&&&& \cr
0 & 12 & 0 & 0 & 0 &
   {9\over 2} & {{11}\over 2} & 3 & {{15}\over 2} & -{3\over 2} & 3 & 4 &
   -{1\over 2} & {{21}\over 2} & {3\over 2} & 6 & {3\over 2} & 9 & 0 &
   {9\over 2} & 0 & 1 & {{15}\over 2} & 3 & -{1\over 2} & 6 & {3\over 2} &
   {9\over 2} & 0 & 3 & {3\over 2} \cr
&&&&&&&&&&&&&&&&&&&&&&&&&&&&&& \cr
1 & 1 & 0 & 0 & 4 & 3 & 0 & -1 & 2 & 0
    & 3 & 0 & 1 & 1 & -1 & 2 & 3 & 1 & -1 & 2 & 3 & 0 & 1 & 2 & 0 & 1 & 2 & 1
    & 2 & 1 & 1 \cr
&&&&&&&&&&&&&&&&&&&&&&&&&&&&&& \cr
1 & 1 & 0 & 2 & 1 & 1 & 1 & 0 & 1 & 2 & 1 & 1 & 1 & 1 & 0
    & 1 & 1 & 1 & 0 & 1 & 1 & 1 & 1 & 1 & 1 & 1 & 1 & 1 & 1 & 1 & 1 \cr
&&&&&&&&&&&&&&&&&&&&&&&&&&&&&& \cr
1 & 9
    & 0 & 0 & 1 & 4 & 4 & 2 & 6 & -1 & 3 & 3 & 0 & 8 & 1 & 5 & 2 & 7 & 0 & 4
    & 1 & 1 & 6 & 3 & 0 & 5 & 2 & 4 & 1 & 3 & 2 \cr
&&&&&&&&&&&&&&&&&&&&&&&&&&&&&& \cr
2 & 2 & 0 & 1 & 2 & 2 & 1
    & 0 & 2 & 1 & 2 & 1 & 1 & 2 & 0 & 2 & 2 & 2 & 0 & 2 & 2 & 1 & 2 & 2 & 1 &
   2 & 2 & 2 & 2 & 2 & 2 \cr
&&&&&&&&&&&&&&&&&&&&&&&&&&&&&& \cr
3 & 3 & 0 & 0 & 3 & 3 & 1 & 0 & 3 & 0 & 3 & 1 & 1
    & 3 & 0 & 3 & 3 & 3 & 0 & 3 & 3 & 1 & 3 & 3 & 1 & 3 & 3 & 3 & 3 & 3 & 3
    \cr &&&&&&&&&&&&&&&&&&&&&&&&&&&&&& \cr
3 & 11 & 0 & 0 & 0 & 4 & 5 & 3 & 7 & -1 & 3 & 4 & 0 & 10 & 2 & 6 & 2
    & 9 & 1 & 5 & 1 & 2 & 8 & 4 & 1 & 7 & 3 & 6 & 2 & 5 & 4 \cr
&&&&&&&&&&&&&&&&&&&&&&&&&&&&&& \cr
4 & 0 & 0 & 0
    & 4 & {5\over 2} & -{1\over 2} & -1 & {3\over 2} & {1\over 2} & 3 & 0 &
   {3\over 2} & {1\over 2} & -{1\over 2} & 2 & {7\over 2} & 1 & 0 & {5\over 2}
    & 4 & 1 & {3\over 2} & 3 & {3\over 2} & 2 & {7\over 2} & {5\over 2} & 4 &
   3 & {7\over 2} \cr
&&&&&&&&&&&&&&&&&&&&&&&&&&&&&& \cr
8 & 0 & 0 & 1 & 2 & 1 & 0 & 0 & 1 & 2 & 2 & 1 & 2 & 1 &
   1 & 2 & 3 & 2 & 2 & 3 & 4 & 3 & 3 & 4 & 4 & 4 & 5 & 5 & 6 & 6 & 7 \cr
&&&&&&&&&&&&&&&&&&&&&&&&&&&&&& \cr
9 & 1
    & 0 & 0 & 3 & 2 & 0 & 0 & 2 & 1 & 3 & 1 & 2 & 2 & 1 & 3 & 4 & 3 & 2 & 4 &
   5 & 3 & 4 & 5 & 4 & 5 & 6 & 6 & 7 & 7 & 8 \cr
&&&&&&&&&&&&&&&&&&&&&&&&&&&&&& \cr
9 & 9 & 0 & 0 & 0 & 3 & 4 & 3
    & 6 & 0 & 3 & 4 & 1 & 9 & 3 & 6 & 3 & 9 & 3 & 6 & 3 & 4 & 9 & 6 & 4 & 9 &
   6 & 9 & 6 & 9 & 9 \cr
&&&&&&&&&&&&&&&&&&&&&&&&&&&&&& \cr
12 & 0 & 0 & 0 & 3 & {3\over 2} & -{1\over 2} & 0 &
   {3\over 2} & {3\over 2} & 3 & 1 & {5\over 2} & {3\over 2} & {3\over 2} & 3
    & {9\over 2} & 3 & 3 & {9\over 2} & 6 & 4 & {9\over 2} & 6 & {{11}\over 2}
    & 6 & {{15}\over 2} & {{15}\over 2} & 9 & 9 & {{21}\over 2} \cr
&&&&&&&&&&&&&&&&&&&&&&&&&&&&&& \cr
16 & 0 & 0
    & 1 & 1 & 0 & 0 & 1 & 1 & 3 & 2 & 2 & 3 & 2 & 3 & 3 & 4 & 4 & 5 & 5 & 6 &
   6 & 6 & 7 & 8 & 8 & 9 & 10 & 11 & 12 & 14 \cr
&&&&&&&&&&&&&&&&&&&&&&&&&&&&&& \cr 16 & 0 & 1 & 0 & 1 & 0 & 0 &
   2 & 1 & 2 & 2 & 2 & 3 & 2 & 4 & 3 & 4 & 4 & 6 & 5 & 6 & 6 & 6 & 7 & 8 & 8
    & 9 & 10 & 11 & 12 & 14 \cr
&&&&&&&&&&&&&&&&&&&&&&&&&&&&&& \cr
24 & 0 & 0 & 1 & 0 & -1 & 0 & 2 & 1 & 4 & 2 &
   3 & 4 & 3 & 5 & 4 & 5 & 6 & 8 & 7 & 8 & 9 & 9 & 10 & 12 & 12 & 13 & 15 & 16
    & 18 & 21 \cr
&&&&&&&&&&&&&&&&&&&&&&&&&&&&&& \cr
24 & 0 & 1 & 0 & 0 & -1 & 0 & 3 & 1 & 3 & 2 & 3 & 4 & 3 & 6
    & 4 & 5 & 6 & 9 & 7 & 8 & 9 & 9 & 10 & 12 & 12 & 13 & 15 & 16 & 18 & 21
    \cr &&&&&&&&&&&&&&&&&&&&&&&&&&&&&& \cr
27 & 3 & 0 & 0 & 0 & 0 & 1 & 3 & 3 & 3 & 3 & 4 & 4 & 6 & 6 & 6 & 6 & 9
    & 9 & 9 & 9 & 10 & 12 & 12 & 13 & 15 & 15 & 18 & 18 & 21 & 24 \cr
&&&&&&&&&&&&&&&&&&&&&&&&&&&&&& \cr
33 & 1
    & 0 & 0 & 0 & -1 & 0 & 3 & 2 & 4 & 3 & 4 & 5 & 5 & 7 & 6 & 7 & 9 & 11 & 10
    & 11 & 12 & 13 & 14 & 16 & 17 & 18 & 21 & 22 & 25 & 29 \cr
&&&&&&&&&&&&&&&&&&&&&&&&&&&&&& \cr
36 & 0 & 0 & 0
    & 0 & -{3\over 2} & -{1\over 2} & 3 & {3\over 2} & {9\over 2} & 3 & 4 &
   {{11}\over 2} & {9\over 2} & {{15}\over 2} & 6 & {{15}\over 2} & 9 & 12 &
   {{21}\over 2} & 12 & 13 & {{27}\over 2} & 15 & {{35}\over 2} & 18 &
   {{39}\over 2} & {{45}\over 2} & 24 & 27 & {{63}\over 2} \cr
&&&&&&&&&&&&&&&&&&&&&&&&&&&&&& \cr  } .
\hspace*{-12mm}
\]}
\\
\centerline{{\bf Table 1:} Pairing matrix for candidates for vertices}
\end{figure}

\noindent
{\bf Example:} The following example is motivated by the non-degenerate
Landau--Ginzburg potential
\beq
	W=x_1^{25}+x_2^8x_1+x_3^3x_5+x_4^3x_2+x_5^3x_1+\l x_2^3x_5^2,\label P
\eeq
to which we assign the matrix
\beq
	A=
\pmatrix{ 25 & 1 & 0 & 0 & 1 \cr 0 & 8 & 0 & 1 & 0 \cr 0 & 0 & 3 & 0 & 0 \cr
   0 & 0 & 0 & 3 & 0 \cr 0 & 0 & 1 & 0 & 3 \cr  }
\eeq
with $q={1\075}( 3,9,17,22,24)$ and $\5q={1\036} ( 1,3,12,12,8 )$.
It is easy to construct all 33 points allowed by the $q$ system (the points in
$\G^4_+$) and the 100 points in $\5\G^4_+$.
%points allowed by the $\5q$ system.
With the help of
\beq
	A^{-1}=
\pmatrix{ {1\over {25}} & -{1\over {200}} & {1\over {225}} & {1\over {600}}
    & -{1\over {75}} \cr 0 & {1\over 8} & 0 & -{1\over {24}} & 0 \cr 0 & 0 &
   {1\over 3} & 0 & 0 \cr 0 & 0 & 0 & {1\over 3} & 0 \cr 0 & 0 & -{1\over 9}
    & 0 & {1\over 3} \cr  }
\eeq
we get the $33\times 100$ matrix of point pairings, which turns out to
have half--integer entries. After eliminating all lines and columns with
less than 4 zeroes we get the pairing matrix for candidates
for vertices shown in table 1.
The first five lines and columns indicate coordinates w.r.t. the
coordinate systems defined by the $q$ and $\5q$ system. All
entries ($i$'th line, $j$'th column) are
$\sum_{k,l=1}^5(\5P^i)_k(A^{-1})^k{}_l(P_j)^l$.
The occurrence of half integers means that we still have a $\BZ_2$ freedom
in choosing sublattices. Eliminating all columns with non--integer entries
corresponds to choosing $\G=\G^4/\BZ_2$ and $\G^*=\5\G^4$, whereas
eliminating all lines with non--integer entries
corresponds to choosing $\G=\G^4$ and $\G^*=\5\G^4/\BZ_2$.
In the first case we would eliminate $P_6$ and $P_7$ which would
result in a first line with only two entries of 0, in contradiction
with our requirement that $\5V^1$ is a vertex of $\D^*$.
Transversality of the polynomial (\ref P) requires the presence of its last
monomial $x_2^3x_5^2$, which is not invariant under this $\IZ_2$ twist
(see the column corresponding to $P_6$ in table 1).
In our context, the $\IZ_2$ twist is forbidden by the requirement that
the vertices of $\5M$ remain vertices of $\D^*$ (dropping this requirement,
the $\IZ_2$ twist may and does lead to reflexive pairs).
In the case $\G=\G^4$ the full matrix of point pairings is a $33\times 52$
matrix.
The convex hulls of the points $P_j$ and $\5P^i$ are polytopes $\D_{\rm max}$
and $\5\D_{\rm max}$, respectively, which are obviously not dual to one
another, as the entries $-1$ show. We can now choose any subset of our
candidates of vertices (containing the vertices in $A$), thus defining
some polytope $\D$. Then we have to eliminate all points of $\5\D_{\rm max}$
which have negative pairings with vertices of $\D$, resulting
in $\G^*\cap\D^*\subseteq\D^*$. Then $\D$ is reflexive if and only if each
of its vertices has pairings of 0 with 4 linearly independent points of
$\G^*\cap\D^*$.
If we keep all points of $\D_{\rm max}$, for example, we have to delete all
lines containing $-1$.
It turns out that this indeed leads to a reflexive pair \cite{ho95}.
In fact, it was checked numerically for all transversal polynomials
that $\D_{\rm max}$ is reflexive \cite{ca95,klun}.
In \cite{wtc} there is also an explicit proof (again, for $n\le 4$)
that $\D_{\rm max}$ is always reflexive even for a larger class of 
weight systems.

If the minimal polytope $M$ is not a simplex
we define a weight system for each of the lower
dimensional simplices $S_\m$ ($\m=1,\cdots,\l$) occurring in corollary 1.
Then we have lattices $\G^k\cong\BZ^k$
and $\5\G^{\5k}\cong\BZ^{\5k}$ and their rational extensions
$\G^k_\BQ\cong\BQ^k$ and $\5\G^{\5k}_\BQ\cong\BQ^{\5k}$, and
we can interpret our coordinate systems as coordinates in $\G^k_\BQ$
and $\5\G^{\5k}_\BQ$.
We get $\l=\5k-n$ equations of the type $\sum \5q^j\5x_j=1$.
Due to the structure given in the lemma, we can solve this system
by successively eliminating the $\5x_{\5k_\m},\;\m=1,\cdots,\l$. Therefore
%Linear independence of the weight systems implies that
these $\5k-n$ equations define an
$n$--dimensional affine hyperplane $\5\G^n_\BQ$ spanned by the $\5V^i$,
which obviously contains $\ipb$ again.
In the same way we also get $k-n$ equations of the type
$\sum q_ix^i=1$ defining an
$n$--dimensional affine hyperplane $\G^n_\BQ$ spanned by the $V_j$.

\section{A classification algorithm}

The classification of dual pairs of reflexive polyhedra can be done
in 3 steps:\\
(1) Classification of possible structures of minimal polytopes,\\
(2) Classification of weight systems,\\
(3) Construction of complete vertex pairing matrices for dual pairs of 
polytopes and choice of a lattice.

Let us first discuss the classification of possible structures of minimal 
polytopes.
With the help of lemma 1 of section 2, it is easy to construct
all possible structures recursively. For a given dimension $n$, one either
has the $n$-dimensional simplex or one has to consider all minimal polytopes
of dimension $n'$ with
$n/2\le n' <n$, add $n-n'+1$ points and consider all possible
structures of $S$ compatible with the lemmata.

For $n=2$ this allows the triangle $V_1V_2V_3$ and the ``1 simplex''
$V_1V_2$ with 2 additional points $V_1', V_2'$, which can only
form another 1 simplex.
In $n=3$ dimensions we can either have a 3 simplex $V_1V_2V_3V_4$
or a 2 dimensional minimal polytope with 2 more points.
The latter case allows the possibilities $S_1=V_1V_2V_3$, $S_2=V_1'V_2'$;
$S_1=V_1V_2V_3$, $S_2=V_1V_2'V_3'$; $S_1=V_1V_2$, $S_2=V_1'V_2'$,
$S_3=V_1''V_2''$.

In $n=4$ dimensions we can have a 4 simplex,
a 3 dimensional minimal polytope with 2 more points or a 2 dimensional
minimal polytope with 3 more points defining a 2 simplex.
The complete list of possible structures is the following: With 5 points
there only is the 4 simplex
$M=S_1=V_1V_2V_3V_4V_5$. With a total of 6 points we have the 4 minimal
configurations							\\[5pt]
\centerline{	\{$S_1=V_1V_2V_3V_4$, $S_2=V_1V_2V_3'V_4'$\},
{}~~~~~~~~~~~~	\{$S_1=V_1V_2V_3V_4$, $S_2=V_1V_2'V_3'$\},}
\centerline{~	\{$S_1=V_1V_2V_3V_4$, $S_2=V_1'V_2'$\},	~~~~~
{}~~~~~~~~~~~~	\{$S_1=V_1V_2V_3$, $S_2=V_1'V_2'V_3'$\}.}	\\[5pt]
With 7 points there are the 3 possibilities			\\[5pt]
\centerline{\{$S_1=V_1V_2V_3$, $S_2=V_1'V_2'$, $S_3=V_1''V_2''$\},}
\centerline{\{$S_1=V_1V_2V_3$, $S_2=V_1V_2'V_3'$, $S_3=V_1''V_2''$\},}
\centerline{\{$S_1=V_1V_2V_3$, $S_2=V_1V_2'V_3'$, $S_3=V_1V_2''V_3''$\},}
								\\[5pt]
and with 8 points we can have only 1 simplices
$S_1=V_1V_2$, $S_2=V_1'V_2'$, $S_3=V_1''V_2''$, $S_4=V_1'''V_2'''$.

The next step in the classification program, namely the 
classification of weight systems, was done in a different
paper \cite{wtc}.
There, weight systems with up to 5 weights and with the property
that $\ip$ is in the interior of $\D_{\rm max}$ were completely
classified.
All weight systems occurring in our scheme (whether alone
or in combination with other weight systems) obviously must
have this ``interior point property''.
The fact that $Q$ consists of hyperplanes carrying facets of $\D$ leads
to another property of weight systems which we may call the
``span property''.
It asserts that the facets of $Q$ must actually be affinely spanned by
points of $\D_{\rm max}$.
According to \cite{wtc}, there are the following weight systems with 
the interior point property:
With two weights, there is only (1/2, 1/2), which also has the span property;
with three weights there are the systems (1/3, 1/3, 1/3), (1/2, 1/4, 1/4) and
(1/2, 1/3, 1/6), which all have the span property as well.
There are 95 systems of four weights (58 of them with the span property),
and there are 184026 systems of five weights (38730 with the span property).

With these informations, there are essentially two ways to construct
all reflexive polyhedra for a given $n$.
We can either pick a specific structure and a combination of weight
systems both for $M$ and $\5M$. 
Then it is easy to write a computer program that finds all $k\times \5k$ 
matrices $A$ that are compatible with such structures, and
we can proceed as in the previous section.
Alternatively, we may give up the symmetry between $\G$ and $\G^*$
%in our point of view 
and simply construct the polyhedron $\D_{\rm max}$
corresponding to some combination of weight systems.
Next, we would consider all of its subpolyhedra $\D$ such that the facets 
of $Q$ are affinely spanned by points of $\D$.
Finally, we must classify all sublattices of $\G^n$ that contain
all the vertices of $\D$ and check for reflexivity of $\D$ w.r.t.
any of these lattices.
In both approaches it is important to calculate and store the 
pairing matrices for the vertices of $\D$ and $\D^*$ because
this is the information required for identifying or distinguishing
polyhedra.

\del
Now consider a structure containing several $q$ and $\5q$ systems.
We pick a specific $q$ system, i.e. a specific set of lines of the
$k\times \5k$ matrix $A$ (but keep all columns).
We may view the $\5k$ columns of the $l\times\5k$ matrix we are looking for
as points in an $l-1$ dimensional affine space defined by the $q$ system.
This space is just a projection of $\G^k_\BQ$, and therefore the $\5k$
points must have $\ip_l$ (the projection of $\ip$) in their interior.
Among these points we can then find
a minimal system again, i.e. we have to solve the classification problem
in $l-1$ dimensions. 
In particular, this approach shows us that when we do
the classification for $n=1,2,3,\cdots$ dimensions, we get new $q$ systems 
only when the minimal polytopes are simplices.

There is also a different approach to the classification of reflexive
polyhedra: After the classification of matrices $A$, we might just keep
the information on the allowed $q$ systems, which is enough to  construct
$\G^n_+$. Then we can simply take all subsets of $\G^n_+$ such that
each coordinate hyperplane is spanned by points in such a subset
and check for reflexivity.
In the following section we will apply this approach to the classification
of reflexive polygons.

\section{The case $n=2$}
\enddel
As an illustraion for our concepts and methods, we will now rederive
the well-known (see, e.g., \cite{ba85,ko90}) classification of reflexive 
polyhedra for $n=2$ in the ``asymmetric'' approach.

In two dimensions there are only two minimal polytopes,
namely the triangle $V_1V_2V_3$ and the parallelogram $V_1V_2V_1'V_2'$.
Thus we either have a single weight system of one of 
the types (1/3, 1/3, 1/3), ~(1/2, 1/4, 1/4),  ~(1/2, 1/3, 1/6), or the
combination of weight systems (1/2, 1/2, 0, 0; 0, 0, 1/2, 1/2).
The points (except $\ip$) allowed in the systems of type $(q_1,q_2,q_3)$
can be arranged as columns of the matrices
\beq
\pmatrix{0&0&0&0&1&2&3&2&1\cr
         0&1&2&3&2&1&0&0&0\cr
         3&2&1&0&0&0&0&1&2\cr},
\eeq
\beq
\pmatrix{0&0&0&0&0&1&2&1\cr
         0&1&2&3&4&2&0&0\cr
         4&3&2&1&0&0&0&2\cr}
\eeq
and
\beq
\pmatrix{0&0&0&0&2&1\cr
         0&1&2&3&0&0\cr
         6&4&2&0&0&3\cr},
\eeq
respectively (see fig. 1).

\BC	\unitlength=2.7pt % \def\putdot#1){\putc#1){\putc(1.5,0){\circle*3}}}
\BP(40,60)(-20,-30)	\thicklines % \linethickness{3pt}
	\putlin(-18,-18,1,0,36)\putlin(-18,-18,1,2,18)\putlin(18,-18,-1,2,18)
	\putdot(-18,-18)\putdot(-6,-18)\putdot(6,-18)\putdot(18,-18)
	\putdot(-12,-6) \putdot(12,-6) \putdot(-6,6) \putdot(6,6) \putdot(0,18)
	\putdot(0,-6)
	\putvec(25,-18,0,1,5)		\putc(28,-14){\small$x_1$}
%	\putvec(20,-22,-2,-1,4)		\putc(14,-22){\small$x_3$}
	\putvec(2,22,2,-1,4)		\putc(8,22){\small$x_2$}
	\putvec(-2,22,-2,-1,4)		\putc(-8,22){\small$x_3$}
			\thinlines
	\putlin(-29,-18,1,0,58) \putlin(-26,-6,1,0,52)
	  \putlin(-20,6,1,0,40)	\putlin(-15,18,1,0,30)
	\putlin(-21,-24,1,2,25)	\putlin(-9,-24,1,2,22)
	  \putlin(3,-24,1,2,16)	\putlin(15,-24,1,2,10)
	\putlin(21,-24,-1,2,25)	\putlin(9,-24,-1,2,22)
	  \putlin(-3,-24,-1,2,16)\putlin(-15,-24,-1,2,10)
	\putnum(-22,-20.5)1	\putnum(-6,-23) 2 	\putnum(6,-23)   3
	\putnum(22,-20.5) 4	\putnum(16,-3.5)5   	\putnum(10,8.5)  6
	\putnum(0,23)     7	\putnum(-10,8.5)8	\putnum(-16,-3.5)9
\EP\hfill
\BP(40,60)(-20,-30)	\thicklines\put(0,-30){\makebox(0,0)[t]
{{\bf
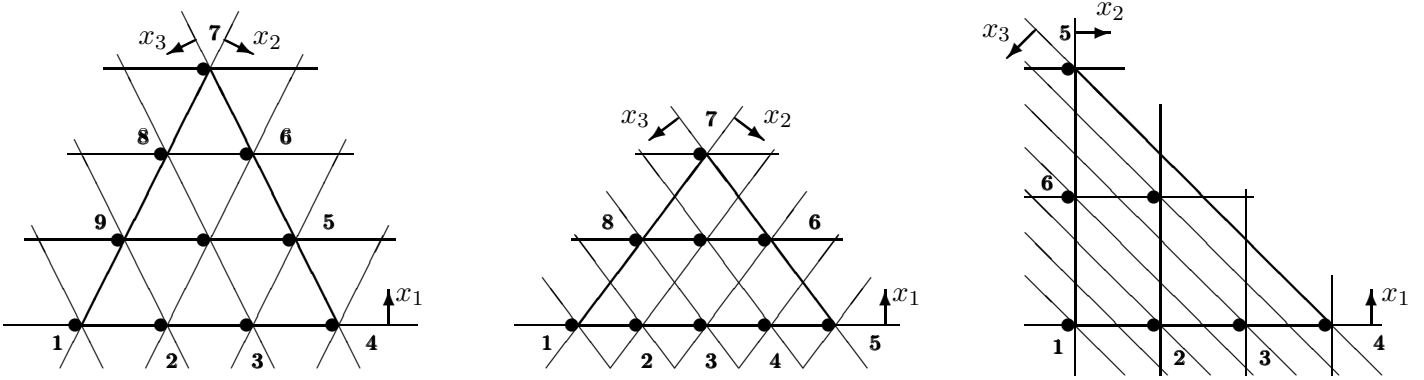
	Fig. 1:} The bounding simplices $Q$ for the weight systems
   $({1\over 3},{1\over 3},{1\over 3})$,  $({1\over 2},{1\over 4},{1\over 4})$
and $({1\over 2},{1\over 3},{1\over 6})$.
}}
	\putlin(-18,-18,1,0,36)\putlin(-18,-18,3,4,18)\putlin(18,-18,-3,4,18)
	\putdot(-18,-18)\putdot(-9,-18)\putdot(0,-18)\putdot(9,-18)
	\putdot(18,-18) \putdot(-9,-6) \putdot(9,-6) \putdot(0,6) \putdot(0,-6)
	\putvec(25,-18,0,1,5)		\putc(28,-14){\small$x_1$}
%	\putvec(20,-22,-2,-1,4)		\putc(14,-22){\small$x_3$}
	\putvec(4,11,4,-3,4)		\putc(10,11){\small$x_2$}
	\putvec(-4,11,-4,-3,4)		\putc(-10,11){\small$x_3$}
			\thinlines
	\putlin(-27,-18,1,0,54) \putlin(-19,-6,1,0,38) \putlin(-10,6,1,0,20)
	\putlin(-22.5,-24,3,4,27.5) \putlin(-13.5,-24,3,4,23)
		\putlin(-4.5,-24,3,4,18.5) \putlin(4.5,-24,3,4,14)
		\putlin(13.5,-24,3,4,9.5)
	\putlin(22.5,-24,-3,4,27.5) \putlin(13.5,-24,-3,4,23)
		\putlin(4.5,-24,-3,4,18.5) \putlin(-4.5,-24,-3,4,14)
		\putlin(-13.5,-24,-3,4,9.5)
	\putnum(-23,-20.5)1	\putnum(-9,-23)    2 	\putnum(0,-23)    3
	\putnum(9,-23)    4	\putnum(23,-20.5)  5	\putnum(14.5,-3.5)6
	\putnum(0,11)     7	\putnum(-14.5,-3.5)8
\EP\hfill
\BP(40,60)(-20,-30)\thicklines
	\putlin(-18,-18,1,0,36)\putlin(-18,-18,0,1,36)\putlin(18,-18,-1,1,36)
	\putdot(-18,-18)\putdot(-6,-18)\putdot(6,-18)\putdot(18,-18)
	\putdot(-18,0) \putdot(-18,18) 		\putdot(-6,0)
	\putvec(23.5,-18,0,1,5)		\putc(27,-14){\small$x_1$}
	\putvec(-18,23,1,0,5)		\putc(-13,26){\small$x_2$}
	\putvec(-23.5,23.5,-1,-1,4)		\putc(-29,23){\small$x_3$}
	\thinlines
	\putlin(-25,-18,1,0,50) \putlin(-25,0,1,0,32) \putlin(-25,18,1,0,14)
	\putlin(-18,-25,0,1,50)	\putlin(-6,-25,0,1,38)
		\putlin(6,-25,0,1,26)	\putlin(18,-25,0,1,14)
	\putlin(-11,-25,-1,1,14) \putlin(-5,-25,-1,1,20)
		\putlin(1,-25,-1,1,26) \putlin(7,-25,-1,1,32)
	\putlin(13,-25,-1,1,38) \putlin(19,-25,-1,1,44)\putlin(25,-25,-1,1,50)
	\putnum(-21,-21)  1	\putnum(-4,-22.5) 2 	\putnum(8,-22.5) 3
	\putnum(24,-20.5) 4	\putnum(-20,23)   5	\putnum(-22.5,2)6
\EP
\EC

In the first case there is a $\BZ_3$ sublattice defined by $x^1=x^2$ mod $3$,
which reduces the set of allowed points to
\beq
\pmatrix{0&0&3\cr
         0&3&0\cr
         3&0&0\cr},
\eeq
and in the second case there is a $\BZ_2$ sublattice defined by $x^2=x^3$ mod
$4$, which reduces the allowed points to
\beq
\pmatrix{0&0&0&2\cr
         0&2&4&0\cr
         4&2&0&0\cr}
\eeq
(see fig. 2), whereas in the case of $(1/2,1/3,1/6)$ there is no allowed
sublattice.

\BC	\unitlength=2.7pt % \def\putdot#1){\putc#1){\circle*3}}
\BP(40,60)(-20,-30)	\thicklines\put(43,-30){\makebox(0,0)[t]
{{\bf
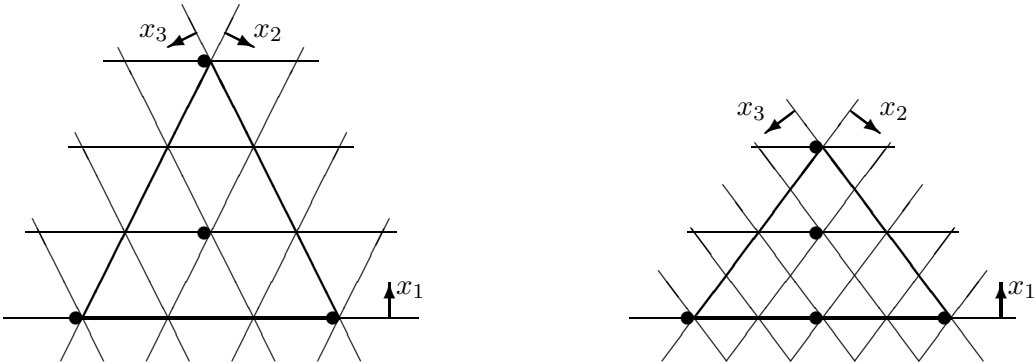
	Fig. 2:} Alternative lattices for the weight systems
   $({1\over 3},{1\over 3},{1\over 3})$ and
     $({1\over 2},{1\over 4},{1\over 4})$.
}}
	\putlin(-18,-18,1,0,36)\putlin(-18,-18,1,2,18)\putlin(18,-18,-1,2,18)
	\putdot(-18,-18)\putdot(18,-18)
	 \putdot(0,18)
	\putdot(0,-6)
	\putvec(25,-18,0,1,5)		\putc(28,-14){\small$x_1$}
%	\putvec(20,-22,-2,-1,4)		\putc(14,-22){\small$x_3$}
	\putvec(2,22,2,-1,4)		\putc(8,22){\small$x_2$}
	\putvec(-2,22,-2,-1,4)		\putc(-8,22){\small$x_3$}
			\thinlines
	\putlin(-29,-18,1,0,58) \putlin(-26,-6,1,0,52)
	  \putlin(-20,6,1,0,40)	\putlin(-15,18,1,0,30)
	\putlin(-21,-24,1,2,25)	\putlin(-9,-24,1,2,22)
	  \putlin(3,-24,1,2,16)	\putlin(15,-24,1,2,10)
	\putlin(21,-24,-1,2,25)	\putlin(9,-24,-1,2,22)
	  \putlin(-3,-24,-1,2,16)\putlin(-15,-24,-1,2,10)
\EP\hspace{42mm}
\BP(40,60)(-20,-30)	\thicklines
	\putlin(-18,-18,1,0,36)\putlin(-18,-18,3,4,18)\putlin(18,-18,-3,4,18)
	\putdot(-18,-18)	\putdot(0,-18)	\putdot(18,-18)
		\putdot(0,6) \putdot(0,-6)
	\putvec(25,-18,0,1,5)		\putc(28,-14){\small$x_1$}
%	\putvec(20,-22,-2,-1,4)		\putc(14,-22){\small$x_3$}
	\putvec(4,11,4,-3,4)		\putc(10,11){\small$x_2$}
	\putvec(-4,11,-4,-3,4)		\putc(-10,11){\small$x_3$}
			\thinlines
	\putlin(-27,-18,1,0,54) \putlin(-19,-6,1,0,38) \putlin(-10,6,1,0,20)
	\putlin(-22.5,-24,3,4,27.5) \putlin(-13.5,-24,3,4,23)
		\putlin(-4.5,-24,3,4,18.5) \putlin(4.5,-24,3,4,14)
		\putlin(13.5,-24,3,4,9.5)
	\putlin(22.5,-24,-3,4,27.5) \putlin(13.5,-24,-3,4,23)
		\putlin(4.5,-24,-3,4,18.5) \putlin(-4.5,-24,-3,4,14)
		\putlin(-13.5,-24,-3,4,9.5)
\EP
\EC

For $(1/3,1/3,1/3)$ we get the triangle $P_1P_4P_7$ twice (both on the original
and the reduced lattice), and in addition we get, on the original lattice,
the polygons
$P_1P_4P_6P_8$, $P_1P_4P_6P_9$, $P_2P_3P_5P_7P_9$, $P_2P_3P_5P_7P_8$,
$P_2P_3P_6P_7P_8$ and $P_2P_3P_5P_6P_8P_9$ (of course, there are more polygons
which are related to the ones given above by the permutation symmetry in the
coordinates).

For $(1/2,1/4,1/4)$ we get the triangle $P_1P_5P_7$ twice (both on the original
and the reduced lattice), and in addition we get the polygons
$P_1P_4P_6P_7$, $P_1P_3P_6P_7$, $P_1P_2P_6P_7$, $P_2P_4P_6P_7P_8$ and
$P_2P_3P_6P_7P_8$.

For $(1/2,1/3,1/6)$ we get the polygons $P_1P_4P_5$, $P_2P_4P_5P_6$ and
$P_3P_4P_5P_6$.

The only case we have not considered so far is the case of two $q$ systems
with $q_1=q_2=1/2$. Allowed points can be encoded by
\beq
\pmatrix{0&0&0&1&2&2&2&1\cr
         2&2&2&1&0&0&0&1\cr
         0&1&2&2&2&1&0&0\cr
         2&1&0&0&0&1&2&2\cr}
\eeq
(see fig. 3). If we drop any of the vertices $P_1,P_3,P_5,P_7$,
we can find a triangle containing $\D$, so we get only two new polygons,
namely $P_1P_3P_5P_7$ on the original lattice and on the sublattice
defined by $x^1=x^3$ mod $2$ (see fig. 4)

\BC	\unitlength=2.7pt % \def\putdot#1){\putc#1){\circle*3}}
\BP(40,60)(-20,-30)	\thicklines\put(0,-30){\makebox(0,0)[t]
{{\bf
	Fig. 3:} $2+2$ full lattice
}}
	\putlin(-18,-18,1,0,36)\putlin(-18,-18,0,1,36)	\putdot(0,0)
	\putlin(18,18,-1,0,36)\putlin(18,18,0,-1,36)
	\putdot(-18,-18)\putdot(18,-18)\putdot(-18,18)\putdot(18,18)
		\putdot(0,18)\putdot(0,-18)\putdot(18,0)\putdot(-18,0)
			\thinlines
	\putlin(-25,-18,1,0,50) \putlin(-25,18,1,0,50) \putlin(-25,0,1,0,50)
	\putlin(-18,-25,0,1,50) \putlin(18,-25,0,1,50) \putlin(0,-25,0,1,50)
	\putvec(-18,-23,1,0,5)	\putc(-10,-23){\small$x_3$}
	\putvec(23,-18,0,1,5)	\putc(27,-14) {\small$x_1$}
	\putvec(18,23,-1,0,5)	\putc(10,23){\small$x_4$}
	\putvec(-23,18,0,-1,5)	\putc(-27,14) {\small$x_2$}
\EP\hspace{42mm}
\BP(40,60)(-20,-30)	\thicklines\put(0,-30){\makebox(0,0)[t]
{{\bf
	Fig. 4:} $2+2$ alternative lattice
}}
	\putlin(-18,-18,1,0,36)\putlin(-18,-18,0,1,36)	\putdot(0,0)
	\putlin(18,18,-1,0,36)\putlin(18,18,0,-1,36)
	\putdot(-18,-18)\putdot(18,-18)\putdot(-18,18)\putdot(18,18)
			\thinlines
	\putlin(-25,-18,1,0,50) \putlin(-25,18,1,0,50) \putlin(-25,0,1,0,50)
	\putlin(-18,-25,0,1,50) \putlin(18,-25,0,1,50) \putlin(0,-25,0,1,50)
	\putvec(-18,-23,1,0,5)	\putc(-10,-23){\small$x_3$}
	\putvec(23,-18,0,1,5)	\putc(27,-14) {\small$x_1$}
	\putvec(18,23,-1,0,5)	\putc(10,23){\small$x_4$}
	\putvec(-23,18,0,-1,5)	\putc(-27,14) {\small$x_2$}
\EP
\EC

We have constructed some polygons more than once. For example,
$P_1P_2P_6P_7$ in the $(1/2,1/4,1/4)$ system is (up to a reflection)
equivalent to $P_2P_4P_5P_6$ in the $(1/2,1/3,1/6)$ system.
Here this can be seen by inspection.
In our approach this redundancy will be sorted out when we bring the complete
pairing
matrices into a normal form by permutations of columns and lines.
Taking this into account, we arrive at the known 16 reflexive polygons
\cite{ba85,ko90}.

\bigskip

{\it Acknowledgements.}  We would like to thank Victor V. Batyrev,
Albrecht Klemm and \hbox{Mahmoud}
Nikbakht Tehrani for helpful discussions. This work is supported
in part by the {\it Austrian Science Foundation} (FWF) under grant number
P10641-PHY.

%%%%%%%%%%%%%%%%%%%%%%%%	References	%%%%%%%%%%%%%%%%%%%%%%%%

%\newpage

\def\LLab#1{\BP(0,0)\unitlength=1mm\put(-15,0){\makebox(0,0)[br]{\small#1}}\EP}
\def\ifundefined#1{\expandafter\ifx\csname#1\endcsname\relax}

% http://www.objectspace.com	STL<ToolKit> - The Standard Template Library

							\ifundefined{draftmode}

\end{document}